# Precision requirements and innovative manufacturing for ultrahigh precision laser interferometry of gravitational-wave astronomy

Wei-Tou Ni*[a], Sen Han[a], Tao Jin[a]

[a] *School of Optical-Electrical and Computer Engineering, University of Shanghai for Science and Technology, 516 Jun Gong Rd., Shanghai, China 200093*

## ABSTRACT

With the LIGO announcement of the first direct detection of gravitational waves (GWs), the GW Astronomy was formally ushered into our age. After one-hundred years of theoretical investigation and fifty years of experimental endeavor, this is a historical landmark not just for physics and astronomy, but also for industry and manufacturing. The challenge and opportunity for industry is precision and innovative manufacturing in large size – production of large and homogeneous optical components, optical diagnosis of large components, high reflectance dielectric coating on large mirrors, manufacturing of components for ultrahigh vacuum of large volume, manufacturing of high attenuating vibration isolation system, production of high-power high-stability single-frequency lasers, production of high-resolution positioning systems etc. In this talk, we address the requirements and methods to satisfy these requirements. Optical diagnosis of large optical components requires large phase-shifting interferometer; the 1.06 μm Phase Shifting Interferometer for testing LIGO optics and the recently built 24" phase-shifting Interferometer in Chengdu, China are examples. High quality mirrors are crucial for laser interferometric GW detection, so as for ring laser gyroscope, high precision laser stabilization via optical cavities, quantum optomechanics, cavity quantum electrodynamics and vacuum birefringence measurement. There are stringent requirements on the substrate materials and coating methods. For cryogenic GW interferometer, appropriate coating on sapphire or silicon are required for good thermal and homogeneity properties. Large ultrahigh vacuum components and high attenuating vibration system together with an efficient metrology system are required and will be addressed. For space interferometry, drag-free technology and weak-light manipulation technology are must. Drag-free technology is well-developed. Weak-light phase locking is demonstrated in the laboratories while weak-light manipulation technology still needs developments.

**Keywords:** metrological requirements, innovative manufacturing, ultrahigh precision laser interferometry, gravitational-wave astronomy

## 1. INTRODUCTION

One hundred years ago, Einstein predicted gravitational waves (GWs) in general relativity (GR) soon after the completion of GR in 1915.[1,2] He thought that GW radiation is, "in all imaginable cases, a practically vanishing value". Indeed, at that time possible expected source strengths and the detection capability had a huge gap. However, with the great strides in the advance of astronomy and astrophysics and in the development of technology, this gap is bridged.

### 1.1 The observation and technology gap 100 years ago

White dwarf was discovered in 1910 with its density soon estimated. Now we understand that GWs from white dwarf binaries in our Galaxy form a stochastic GW background ("confusion limit")[3] for space (low frequency)[4] GW detection in GR. The characteristic strain for confusion limit is about $10^{-20}$ in 0.1-1 mHz band. As to individual sources, some can have characteristic strain around this level for frequency 1-3 mHz in low-frequency band. One hundred year ago, the sensitivity of astrometric observation through the atmosphere around this band is about 1 arcsec. This means the strain sensitivity to GW detection is about $10^{-5}$; 15 orders away from the required sensitivity. The first artificial satellite Sputnik was launched in 1957. The technological demonstration mission LISA Pathfinder was launched on 3 December, 2015. This mission basically demonstrated the drag-free requirement of LISA GW space mission concept;[5] the major issue in the technological gap 15 orders of magnitude is successfully abridged during last hundred years. This drag-free technology should also be readily applicable to other GW space mission proposals.[4] However, at present the space GW missions are only expected to be launched more than a decade later. Weak-light phase locking is demonstrated in laboratories.[6,7] Weak-light manipulation technology still needs developments. In the LIGO discovery of 2 GW events and 1 probable GW candidate, the maximum peak strain intensity is $10^{-21}$; the frequency range is 30-450 Hz.[8,9] Strain gauge in this frequency region could reach $10^{-5}$ with a fast recorder about 100 years ago; thus, the technology gap is 16 orders of magnitudes. Michelson interferometer for Michelson-Morley experiment[10] has a strain ($\Delta l/l$) sensitivity of $5 \times 10^{-10}$ with 0.01 fringe detectability and 11 m path length; however, the appropriate test mass suspension system with fast (30-450 Hz in the high-frequency GW band) white-light observing system is lacking.

*weitouni@163.com; http://astrod.wikispaces.com

Summarizing, the detection gap between astrophysical GW source strength and the technological achievable sensitivity is about 15-16 orders of magnitude in amplitude in the low-frequency and high-frequency GW bands 100 years ago.

**1.2 The gap largely bridged 50 years ago**

The experimental efforts to detect gravitational waves did not begin until Joseph Weber, his post-doc David Zipoy and student Robert Forward started to work on GW experiment from 1958 on. They were pushing the technology for doing experiment at that time.

In 1966, fifty years after Einstein's first paper on GW and fifty years before LIGO announced their first detection of GW, Weber published a paper[11] on the sensitivity of the 1660 Hz Maryland resonant bar detector "Observation of the Thermal Fluctuations of a Gravitational-Wave detector." In the paper, he stated that strains as small as a few parts in $10^{16}$ are observable for a compressional mode of a large cylinder. This is an important cornerstone and narrows the 16 orders of gap of observable GW strength and experimental sensitivity to 6 orders of gap in the GW detection.

**1.3 The start of precision laser interferometry for GW detection**

Experimental work for first laser interferometer was started in Hughes Research Laboratories in 1966.[12,13] In their 1971 paper[13] there is an optical layout of Michelson interferometer for gravitational radiation antenna transducer. In this paper, Moss, Miller and Forward[13] reported on the achievement of photon-noise-limited performance using 80 μW from a single mode Spectra-Physics 119 laser in a modified Michelson interferometer on a vibration isolation table in a quiet room. They used piezoelectric driver on one of the interferometer mirrors to generate 30 fm ($3 \times 10^{-14}$ m) vibrations of known amplitude (Fig. 1 [left]). The measured displacement sensitivity of the system in the kilohertz region was 13 fm Hz$^{-1/2}$, which compares well with the calculated photon noise limit of 10.6 fm Hz$^{-1/2}$ which was the smallest vibrational displacement measured directly with a laser to that date.

In 1972, Forward[14] and his colleagues completed the first laser-interferometer antenna (Fig. 1 [right]) for GW detection in Hughes Research Laboratories, Malibu, California: "The laser interferometer was operated as a detector for gravitational radiation for 150 h during the nights and weekends from the period 4 October through 3 December 1972." Bar antennas were also operated by the Maryland, Glasgow, and Frascati groups during that period. The various bar antenna systems were sensitive to gravitational-radiation strain spectral components with an amplitude of the order of 0.1 fm/m in a narrow band of frequencies about the resonant frequency of the bar. The Malibu wideband interferometer was estimated to have sensitivity of the order of 1-10 fm/m for the total of the gravitational-radiation strain spectral components in the band from 1-20 kHz. No significant correlations between the Malibu interferometer output and any of the bar events or coincidences were observed.

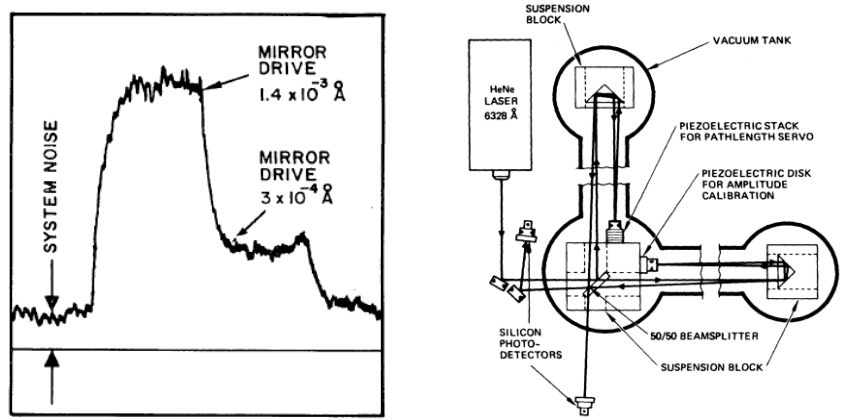

Fig. 1. (left) Interferometer system noise measurement at 5 kHz of Moss, Miller and Forward [13]; (right) Schematic of Malibu Laser Interferometer GW Antenna (from Forward [14]).

In 1972, Weiss[15] wrote a report on "electromagnetically coupled broadband gravitational antenna" in Quarterly Progress Report, Research Laboratory of Electronics (RLE) of MIT. This report was to become a classical source document for large-scale laser interferometry for GW detection. It contained the motivation and the proposed antenna design together with a rather thorough discussion of most fundamental noise sources in the interferometer. The fundamental noise sources discussed include:

       a. Amplitude noise in the laser output power;
       b. Laser phase noise or frequency instability;
       c. Mechanical thermal noise in the antenna;
       d. Radiation-pressure noise from laser light;
       e. Seismic noise;
       f. Thermal-gradient noise;
       g. Cosmic-ray noise;
       h. Gravitational-gradient noise;
       i. Electric field and magnetic field noise.

GW detection requires that the end mirrors of interferometers be free from any nongravitational interactions, i.e. be inertial, in the detection direction. For earth-based interferometer, the end mirrors have to be suspended independently, or floating horizontally. The long list of laser interferometers with independently suspended mirrors also includes the one-arm interferometers with independently suspended mirrors for QED vacuum birefringence experiment Q & A. It measures the birefringence (the optical path length difference for two different polarizations in the same cavity) of vacuum due to magnetic field modulation. These interferometers include Q & A 3.5 m interferometer[16] (1993-2008, Hsinchu), Q & A 7 m interferometer[17] (2008-2015, Hsinchu) and Q & A 9 m interferometer (2016 start, Shanghai).

Major detection development efforts in the high frequency band (10 Hz – 100 k Hz) are in the long arm laser interferometers. The TAMA 300 m arm length interferometer,[18] the GEO 600 m interferometer,[19] and the kilometer size laser-interferometric GW detectors --- LIGO[20] (two 4 km arm length, one 2 km arm length) and VIRGO[21] (3 km arm length) all achieved their original sensitivity goals basically. TAMA 300 and GEO 600 are the long prototypes. LIGO and VIRGO are the first generation GW detectors. Around the frequency 100 Hz, the LIGO and Virgo sensitivities are both in the level of $10^{-23}$ (Hz)$^{-1/2}$. These interferometers paved the road for second generation GW detectors – aLIGO, aVirgo and KAGRA. aLIGO[22] and aVIRGO[23] are upgraded from (initial) LIGO and (initial) VIRGO. KAGRA/LCGT[24] is a 3 km cryogenic underground interferometer started construction in 2010 based on the experience of TAMA 300 m interferometer and CLIO[25] 100 m cryogenic underground interferometer.

The discovery of aLIGO[8,9] with outlooks for future observations including the third generation detectors will be discussed in Subsection 2.1.

## 1.4  Space laser interferometry for GW detection

Doppler tracking of spacecraft is also interferometry and can be used to constrain (or detect) the level of low-frequency GWs.[26] The separated test masses of this GW detector are the Doppler tracking radio antenna on Earth and a distant spacecraft. Doppler tracking measures relative distance-change. Estabrook and Walquist analyzed[26] the effect of GWs passing through the line of sight of spacecraft on the Doppler tracking frequency measurements (see also Ref. 27). From these measurements, GWs can be detected or constrained. The most recent measurements came from the Cassini spacecraft Doppler tracking (CSDT). Armstrong, Iess, Tortora, and Bertotti[28] used precision Doppler tracking of the Cassini spacecraft during its 2001–2002 solar opposition to derive improved observational limits on an isotropic background of low-frequency gravitational waves. They used the Cassini multilink radio system and an advanced tropospheric calibration system to remove the effects of leading noises — plasma and tropospheric scintillation to a level below the other noises. The resulting data were used to construct upper limits on the strength of an isotropic background in the 1 μHz to 1 mHz band.[28]

Space laser interferometers for GW detection hold the most promise with high signal-to-noise ratio. LISA[29] (Laser Interferometer Space Antenna) is aimed at detection of $10^{-4}$ to 1 Hz GWs with a strain sensitivity of $4 \times 10^{-21}$/(Hz)$^{1/2}$ at 1 mHz. See Table 1 for a compilation of GW mission proposals. There are abundant sources for eLISA/LISA, ASTROD, ASTROD-GW and Earth-orbiting missions: (i) In our Galaxy: galactic binaries (neutron stars, white dwarfs, etc.); (ii) Extra-galactic targets: supermassive black hole binaries, supermassive black hole formation; and (iii) Cosmic GW background. A date of launch of eLISA or substitute mission was set around 2034.[46]

LISA Pathfinder was launched on December 3, 2015 and successfully tested the drag-free technology to satisfy not just the requirement of LISA Pathfinder, but also the requirement of LISA.[5] The success paved the road for all the space mission proposal in the table. And we do anticipate the possibility of an earlier launch date for eLISA (or a substitute mission) and possible earlier flight of other missions.

Table 1. A Compilation of GW Mission Proposals

| Mission Concept | S/C Configuration | Arm length | Orbit Period | S/C # |
|---|---|---|---|---|
| *Solar-Orbit GW Mission Proposals* | | | | |
| LISA[29] | Earth-like solar orbits with 20° lag | 5 Gm | 1 year | 3 |
| eLISA[30] | Earth-like solar orbits with 10° lag | 1 Gm | 1 year | 3 |
| ASTROD-GW[31] | Near Sun-Earth L3, L4, L5 points | 260 Gm | 1 year | 3 |
| Big Bang Observer[32] | Earth-like solar orbits | 0.05 Gm | 1 year | 12 |
| DECIGO[33] | Earth-like solar orbits | 0.001 Gm | 1 year | 12 |
| ALIA[34] | Earth-like solar orbits | 0.5 Gm | 1 year | 3 |
| Taiji (ALIA-descope)[35] | Earth-like solar orbits | 3 Gm | 1 year | 3 |
| Super-ASTROD[36] | Near Sun-Jupiter L3, L4, L5 points (3 S/C), Jupiter-like solar orbit(s)(1-2 S/C) | 1300 Gm | 11 year | 4 or 5 |
| *Earth-Orbit GW Mission Proposals* | | | | |
| OMEGA[37] | 0.6 Gm height orbit | 1 Gm | 53.2 days | 6 |
| gLISA/GEOGRAWI[38-40] | Geostationary orbit | 0.073 Gm | 24 hours | 3 |
| GADFLI[41] | Geostationary orbit | 0.073 Gm | 24 hours | 3 |
| TIANQIN[42] | 0.057 Gm height orbit | 0.11 Gm | 44 hours | 3 |
| ASTROD-EM[43,44] | Near Earth-Moon L3, L4, L5 points | 0.66 Gm | 27.3 days | 3 |
| LAGRANGE[45] | Near Earth-Moon L3, L4, L5 points | 0.66 Gm | 27.3 days | 3 |

**1.5 Scope, Precision and Innovation**

In section 2, we project to the future of GW astronomy from the current known aims and current detected events together with the pace of current development of GW detectors. In Section 3, we discuss the precision requirements of Earth-based laser interferometry and space-borne laser interferometry. In section 4, we discuss the experimental efforts and innovative manufacturing to meet these requirements.

## 2. SCOPE OF GW ASTRONOMY

**2.1 Goals of GW astronomy**

The goals of GW measurements are two fold – to do GW astronomy and to explore the fundamental physics related to gravitational phenomena. To do GW astronomy and to explore fundamental physics, every frequency band is being explored. Table 2 summarizes GW frequency classification according to detection methods together with GW sources.[47,48,4]

**2.2 The future of Earth-based interferometry for GW detection**

After more than fifty years of experimental efforts in search for GWs since Weber, the laser-interferometric GW detector aLIGO finally has enough sensitivity to have a good chance to detect GW directly. The aLIGO noise (Fig. 2) reached below $10^{-23}$ Hz$^{-1/2}$ in September 2016 and was able to clearly detect GW strain $10^{-21}$ of duration 30 ms or more around the 100 Hz frequency band.[8]

From the LIGO O1 observation run, there are two GW events with significance greater than 5.3 σ and one likely GW candidate with significance 1.7 σ in 130-days observation.[8,9] These detection events/candidate occurred at September 14, October 12 and December 26 of 2015 with luminosity distance about 420 Mpc, 1000 Mpc and 440 Mpc respectively. Their radiated energies are the equivalent energies of 3, 1.0 and 1.5 solar mass respectively. All events had peak

Table 2. Frequency Classification of Gravitational Waves, GW sources and detection methods.[47,48,4]

| Frequency band | GW sources / Possible GW sources | Detection method |
|---|---|---|
| Ultrahigh frequency band: above 1 THz | Discrete sources, Cosmological sources, Braneworld Kaluza-Klein (KK) mode radiation, Plasma instabilities | Terahertz resonators, optical resonators, and magnetic conversion detectors |
| Very high frequency band: 100 kHz – 1 THz | Discrete sources, Cosmological sources, Braneworld Kaluza-Klein (KK) mode radiation, Plasma instabilities | Microwave resonator/wave guide detectors, laser interferometers and Gaussian beam detectors |
| High frequency band (audio band)*: 10 Hz – 100 kHz | Conpact binaries [NS (Neutron Star)-NS, NS-BH (Black Hole), BH-BH], Supernovae | Low-temperature resonators and Earth-based laser-interferometric detectors |
| Middle frequency band: 0.1 Hz – 10 Hz | Intermediate mass black hole binaries, massive star (population III star) collapses | Space laser-interferometric detectors of arm length 1,000 km − 60,000 km |
| Low frequency band (milli-Hz band)†: 100 nHz – 0.1 Hz | Massive black hole binaries, Extreme mass ratio inspirals (EMRIs), Compact binaries | Space laser-interferometric detectors of arm length longer than 60,000 km |
| Very low frequency band (nano-Hz band): 300 pHz – 100 nHz | Supermassive black hole binary (SMBHB) coalescences, Stochastic GW background from SMBHB coalescences | Pulsar timing arrays (PTAs) |
| Ultralow frequency band: 10 fHz – 300 pHz | Inflationary/primordial GW background, Stochastic GW background | Astrometry of quasar proper motions |
| Extremely low (Hubble) frequency band (cosmological band): 1 aHz – 10 fHz | Inflationary/primordial GW background | Cosmic microwave background experiments |
| Beyond Hubble-frequency band: below 1 aHz | Inflationary/primordial GW background | Through the verifications of inflationary/primordial cosmological models |

*The range of audio band normally goes only to 10 kHz.
†The range of milli-Hz band is 0.1 mHz to 100 mHz.

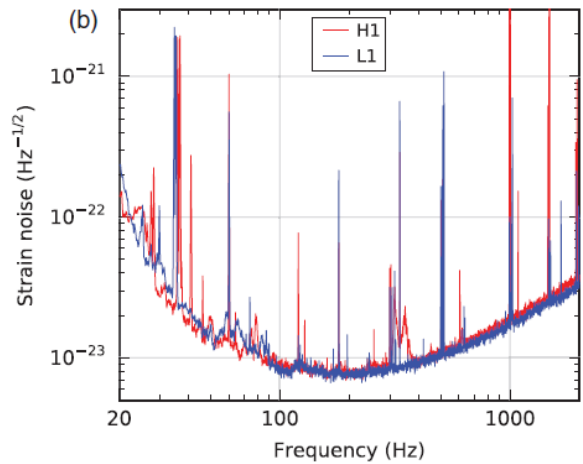

Fig. 2. The instrument noise for each of the two LIGO detectors at Hanford, WA (H1) and Livingston, LA (L1) near the time of the signal detection of GW150914; this is an amplitude spectral density, expressed in terms of equivalent gravitational-wave strain amplitude. The sensitivity is limited by photon shot noise at frequencies above 150 Hz, and by a superposition of other noise sources at lower frequencies [49]. Narrow-band features include calibration lines (33-38, 330, and 1080 Hz), vibrational modes of suspension fibers (500 Hz and harmonics), and 60 Hz electric power grid harmonics. (Figure and caption from Fig. 3 of [8])

luminosity around $3 \times 10^{56}$ erg s$^{-1}$. The peak intensity of the first detected event GW150914 reached a strain of $10^{-21}$ with frequency around 100 Hz. These together with some other characteristics deduced from LIGO O1 GW observations are listed in Table 3.

Table 3. Characteristics of two GW events and one GW candidate deduced from LIGO O1 GW observations.[8,9]

| Event | GW150914 | GW151226 | LVT151012 (candidate) |
|---|---|---|---|
| Signal-to-noise ratio $\rho$ | 23.7 | 13.0 | 9.7 |
| Significance | $> 5.3\ \sigma$ | $> 5.3\ \sigma$ | $1.7\ \sigma$ |
| Primary mass $m^{source}_1/M_\odot$ | $36.2^{+5.2}_{-3.8}$ | $14.2^{+8.3}_{-3.7}$ | $23^{+18}_{-6}$ |
| Secondary mass $m^{source}_2/M_\odot$ | $29.1^{+3.7}_{-4.4}$ | $7.5^{+2.3}_{-2.3}$ | $13^{+4}_{-5}$ |
| Effective inspiral spin $\chi_{eff}$ | $-0.06^{+0.14}_{-0.14}$ | $0.21^{+0.20}_{-0.10}$ | $0.0^{+0.3}_{-0.2}$ |
| Final mass $M^{source}_f/M_\odot$ | $62.3^{+3.7}_{-3.1}$ | $20.8^{+6.1}_{-1.7}$ | $35^{+14}_{-4}$ |
| Final spin $a_f$ | $0.68^{+0.05}_{-0.06}$ | $0.74^{+0.06}_{-0.06}$ | $0.66^{+0.09}_{-0.10}$ |
| Radiated energy $E_{rad}/(M_\odot c^2)$ | $3.0^{+0.5}_{-0.4}$ | $1.0^{+0.1}_{-0.2}$ | $1.5^{+0.3}_{-0.4}$ |
| Peak luminosity $l_{peak}/(\text{erg s}^{-1})$ | $3.6^{+0.5}_{-0.4} \times 10^{56}$ | $3.3^{+0.8}_{-1.6} \times 10^{56}$ | $3.1^{+0.8}_{-1.8} \times 10^{56}$ |
| Luminosity distance $D_L$/Mpc | $420^{+150}_{-180}$ | $440^{+180}_{-190}$ | $1000^{+500}_{-500}$ |
| Source redshift $z$ | $0.09^{+0.03}_{-0.04}$ | $0.09^{+0.03}_{-0.04}$ | $0.2^{+0.09}_{-0.09}$ |

The O1 observation period spanned 130 days from September 12, 2015 to January 19, 2016. If we simply scale this to one year observation, we would have chance to observe five GW events with significance greater than 5.3 σ. When the advanced LIGO goal sensitivity is reached, there would be a 3 fold improvement and 27 fold reach in volume (still basically local universe); hence more than 100 GW events per year with significance greater than 5.3 σ. For the third-generation Earth-based laser interferometer for GW detection (like ET), there would be another tenfold increase in sensitivity and 1000-fold reach in volume; hence 100 k GW events per year. These would be a lot for studying astronomy and cosmology.

**2.3 The future of space-borne interferometry for GW detection**

The first public proposal on space interferometers for GW detection was presented at the Second International Conference on Precision Measurement and Fundamental Constants (PMFC-II), 8–12 June 1981, in Gaithersburg [50,51]. In this seminal proposal, Faller and Bender raised possible GW mission concepts in space using laser interferometry. Two basic ingredients were addressed — drag-free navigation for the reduction of perturbing forces on the spacecraft (S/C) and laser interferometry for the sensitivity of measurement. LISA-like S/C orbit formation was reached in 1985 in the proposal Laser Antenna for Gravitational-radiation Observation in Space (LAGOS).[52]

There are plenty of GW sources in the low-frequency region (100 nHz to 100 mHz). They even form a confusion limit of galactic background for space detection as we discussed in the introduction. Some GW sources may have signal to noise ratio in the range of $10^4$ to $10^5$. The science goals for space missions are the detection of GWs from (i) Supermassive Black Holes; (ii) Extreme-Mass-Ratio Black Hole Inspirals; (iii) Intermediate-Mass Black Holes; (iv) Galactic Compact Binaries and (v) Relic GW Background.

With LISA Pathfinder successfully tested the drag-free technology to satisfy not just the requirement of LISA Pathfinder, but also[5] basically the requirement of LISA and the first detection of GWs by aLIGO, we do anticipate the possibility of an earlier launch date for eLISA (or a substitute mission) and likely earlier flight of other missions.

**2.4 Various people working for GW astronomy**

Considering multi-disciplinary nature of GW astronomy, there will be three kinds of GW researchers in future:

(i) Experimentalists (Experimental Astronomers/Engineers/Physicists), working on detectors and data processing;
(ii) Multi-Messenger Astronomers, working on astrophysics;
(iii) Theoretical Physicists/Cosmologists, working on fundamental physics and theoretical cosmology.

- Budget: grow up to 20 % - 30 % of Astronomy Budget

## 2.5 Other bands of GW detection

For other bands of GW detection, see Kuroda, Ni and Pan [4].

## 3. PRECISION REQUIREMENTS

### 3.1 Precision requirements of Earth-based laser interferometry for GW detection

Since the design sensitivities of aLIGO, aVIRGO and KAGRA/LCGT are similar and KAGRA/LCGT design has some of the 3rd generation characteristics, we use KAGRA/LCGT requirements for illustration. The left part of Fig. 3 shows the design sensitivity of broad-band resonant side-band extraction (BRSE) configuration.[53] The right part of Fig. 3 shows the design sensitivity of detuned resonant side-band extraction (DRSE) configuration.[53] There are 5 length degrees of freedom of KAGRA interferometer to do feed back control: DARM (Differential length change of the arm cavities), CARM Common length change of the arm cavities), MICH (Michelson degree of freedom), PRCL (Power-recycling cavity length), and SRCL (Signal-recycling cavity length). For an optimization of inspiral range (IR), the following parameters related to quantum noises of the KAGRA interferometer are chosen as: Arm cavity finesse, 1530; ITM (Input Test Mass) reflectivity, 99.6%; ITM loss < 50 ppm; PRM (Power Recycling Mirror) reflectivity, 90%; ETM (End Test Mass) loss < 50 ppm; Homodyne angle 132°; SRM (Signal Recycling Mirror) reflectivity, 85%; Input laser power, 78 W; Detuning angle, 3.5°; DRSE IR, 237 Mpc; BRSE IR; 217 Mpc. The length and frequency parameters are chosen as: Arm cavity length, 3000 m; PRC length, 66.591 m; SRC length, 66.591 m; Michelson asymmetry, 3.30 m; RF sideband f1 frequency, 16.881 MHz; RF sideband f2 frequency, 45.016 MHz; MC length, 26.639 m. For a detailed account, see [53].

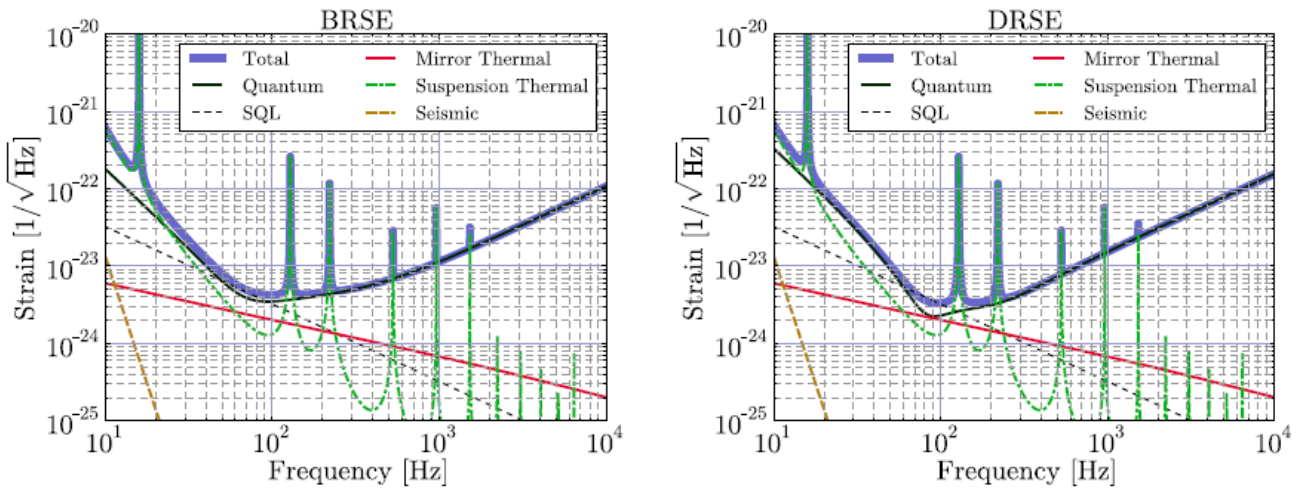

Fig. 3. Design sensitivity of KAGRA. DRSE, shown in the right-hand side figure is more sensitive at frequencies of less than 500Hz, while BRSE in the left-hand side figure is better at higher frequencies. (from [53])

### 3.2 Precision requirements of space-borne laser interferometry for GW detection

The drag-free requirement of all first-generation mission proposal are all similar to LISA and has been largely achieved by LISA Pathfinder as shown in Fig. 4. The weak-light phase locking is crucial for long-distance space interferometry. For LISA of arm length of 5 Gm (million km) the weak-light phase locking requirement is for 70 pW laser light to phase-lock with an onboard laser oscillator. For ASTROD-GW arm length of 260 Gm (1.73 AU) the weak-light phase locking requirement is for 100 fW laser light to lock with an onboard laser oscillator. Weak-light phase locking for 2 pW laser light to 200 μW local oscillator is demonstrated in our laboratory in Tsing Hua U.[6] Dick *et al.*[7] from their phase-locking experiment showed a PLL (Phase Locked Loop) phase-slip rate below one cycle slip per second at powers as low as 40 fW in JPL.

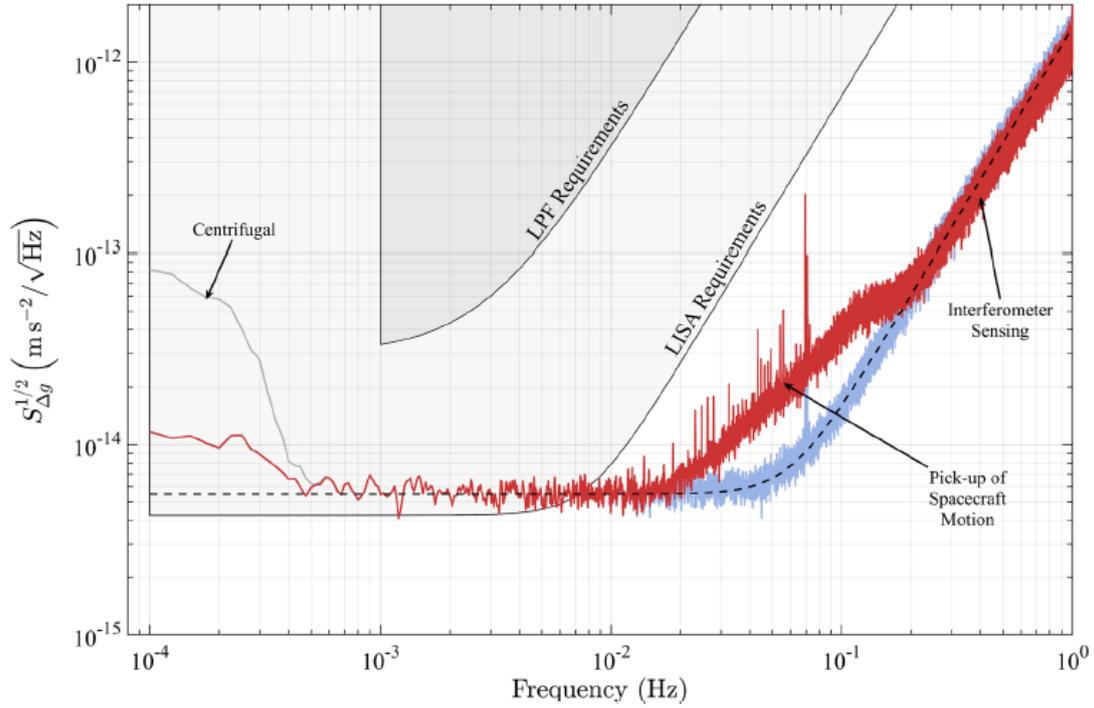

Fig. 4. Shaded areas: LISA and LISA Pathfinder requirements for Δg. The LISA single test-mass acceleration requirement has been multiplied by $2^{1/2}$ to be presented here as a differential acceleration.[5]

## 4. INNOVATIVE MANUFACTURING

Here we review and address some issues on innovative manufacturing.

### 4.1 Production of large and homogeneous optical components

For KAGRA, production of large sapphire mirrors with 220 mm/250 mm with the required homogeneity and absorption for cryogenic use at 20 K is still an issue. Although large boule of 100 kg could be produced, the quality is still an issue which needs more testing and innovation.

### 4.2 Optical diagnosis of large optical components

The first generation km-sized GW detectors require unprecedented sensitivity to relative movement between the optics in the interferometer arms. The final design consists of optics of extremely high quality with apertures of up to 250 mm. These must be characterized with a high degree of accuracy and precision. Small errors, which normally do not affect the overall acceptance of a test piece, must be analyzed and corrected to ensure the proper parts characterization. Veeco Metrology has successfully designed and built an interferometer to measure the resonator optics.[54]

In order to ensure proper functioning of the LIGO project the resonator optics must be characterized with high accuracy. The instrument specifications called for Fizeau type interferometer operating at wavelength of 1.06 μm (the wavelength of the LIGO interferometers) with the capability of measuring the ROC (Radius Of Curvature) with errors below 3% for radii in the range of 5.5 km to 14.5 km. The interferometer must have a clear aperture of 150 mm and must be able to measure features from 100 mm down to 1 mm on optics with from 4% to 99.9% reflectance. In addition to accurately measuring the radius of curvature, the residual focus and the astigmatism must be accurate to λ/100 P-V. These coefficients are determined from Zernike coefficients using the full-aperture measurement. The residual RMS error after removing the focus and the astigmatism terms (along with the tilt and the piston), must be less than λ/l000. Accurate values of pixel spacing and distortion of the imaging system must be determined for different instrument magnifications.

The retrace error, which refers to the calculated OPD (optical path difference) map difference between null and n-fringe cavities needs to be less than 6 nm P-V for 4 fringes of tilt with an optical flat forming the cavity. The long ROC of the LIGO optics necessitates this requirement, as testing against flat surfaces precludes being able to null the fringe pattern. Ray tracing can model the retrace error, but the retrace error needs to be measured and subtracted if necessary.[54]

With these requirements, Veeco Metrology chose to modify the design of the WYKO 6000 Laser Fizeau Interferometer to accommodate the stringent requirements of the LIGO metrology system by paying special attention to collimator, the focusing and imaging lenses, and system alignment. For a detailed account the more challenging aspects of both the optical design and the alignment, please see Ref. [54].

Even larger optics will be needed for 10-50 km laser interferometers for GW detection. It would be good to give thoughts to larger diagnosis instruments. Modification of the recently built 24" phase-shifting Interferometer in Chengdu, China would be a choice.

### 4.3 High reflectance dielectric coating on large mirrors

This needs is largely achieved by industry now. The industry should be well prepared for even larger mirror for future generations.

### 4.4 Manufacturing of components for ultrahigh vacuum of large volume

For the initial LIGO, about 1/3 of the budget is for land acquisition, 1/3 for vacuum system and 1/3 for everything else. Hence manufacturing of components for ultrahigh vacuum of large volume is a large business. To make the wall of vacuum pipe thinner but reliable, one need inventions. Wall with helical ridges are adopted. Pitch need to be designed optimally. To suppress fluctuations and absorption due to residual gases, ultrahigh vacuum is needed and the water vapor pressure needs to be suppressed below $10^{-9}$ torr. This needs baking out. In the case of LIGO, baking out the LIGO Beam Pipe took 2000 amp current for a month. The current power supplies used were left-overs from SSC (Superconducting Super Collider). For 10-50 km laser interferometers, the vacuum requirement would be more stringent due to longer optical paths.

### 4.5 Manufacturing of high attenuating vibration isolation system

High attenuating vibration isolation system to make the suspended mirrors inertial in the measurement direction is one of the two most important technologies for earth-bound GW detection. The other is interferometric sensing. Many of the components and subsystems became commercial already. The percent will increase. As sensitivity becomes higher, requirements on vibration isolation system will be more stringent and innovation will be needed.

### 4.6 Production of high-power high-stability single-frequency lasers

Production of high-power high-stability single frequency lasers up to 200 W becomes commercial. As the heating issues become lessened, higher power will be needed.

### 4.7 Production of high resolution length measurement and positioning systems[55]

In the drag-free control of spacecraft, we need precision metrology sensor to sense the position of the spacecraft relative to the proof mass. For this purpose, there are two choices – capacitance sensor and laser metrology sensor.[55] For a laser metrology sensor system, larger gap is possible. Hence, less local gravitational disturbances are incurred and better accuracy in controlling the deviation from the geodesics can be achieved. Recently we have proposed to use laser metrology and optical methods for performing an equivalence principle test in space.[56] All these needs picometer real-time laser metrology and real-time motion control. For ultra-high precision interferometric measurement, the optical path length in optical device need to be measured and controlled very accurately also.

For laser metrology, we use mid-point cyclic average to minimize nonlinearity error.[57] After fourth cyclic average, the residual nonlinearity error is about 1.5 pm rms. For real-time control, we need real-time measurement. We use PZT's to modulate a small mirror to perform real-time averaging. We used two other sets of PZT's for motion and motion correction, and reached real-time measuring error of 560 pm and real-time control error of 700 pm rms. This real-time control was for a 440 nm run.[57] To reach a longer span, we use the experimental setup shown in Fig. 4 of [55] for study. For data acquisition, data processing and control, we use a VME-bus Heurikon single board computer (Nitro60 with Motorola 68060 CPU) with VxWorks. We use a PI PZT to drive the flexure stage. The laser metrology mirror is mounted on three PZT's for position measurement. This mirror can be modulated by PZT's for mid-point cyclic average to minimize nonlinearity error in real-time. The PI PZT has a length-change range of 12.6 $\mu$m for 100 V. The nonlinearity and hysteresis of the PZT is measured and complied as a look-up table using the laser interferometer system before real-time motion control is to be implemented. For real-time motion control, the D/A voltage to be applied is first calculated using the look-up table. After the initiation, the error signal for the prescribed motion is feedback to the PI PZT as an increment voltage through D/A. The real-time motion control errors for 10 $\mu$m linear and quadratic motions are about 0.9 nm rms and are dominately quantization errors due to 12 bit DAC. The rms error after 16-point average is 0.21 nm. For longer motions and/or better precisions, more stages could implemented.

High precision optical missions enable us to study fundamental problems in physics and astrophysics. These missions present high-technology challenges which demand matching R & D's. These R & D's will benefit optical space communications and find important Earth-bound applications.

### 4.8 Drag-free technology

Drag-free technology has been achieved for LISA requirement by LISA Pathfinder.[5] Its commercialization or partial commercialization looks a good way to do business.

### 4.9 Weak-light phase locking and manipulation technology

Weak-light phase locking is crucial for long-distance space interferometry and for CW laser space communication. For LISA of arm length of 5 Gm (million km) the weak-light phase locking requirement is for 70 pW laser light to phase-lock with an onboard laser oscillator. For ASTROD-GW arm length of 260 Gm (1.73 AU) the weak-light phase locking requirement is for 100 fW laser light to lock with an onboard laser oscillator. Weak-light phase locking for 2 pW laser light to 200 µW local oscillator is demonstrated in our laboratory in Tsing Hua U.[6] Dick *et al.*[7] from their phase-locking experiment showed a PLL (Phase Locked Loop) phase-slip rate below one cycle slip per second at powers as low as 40 femtowatts (fW). This phase-slip rate corresponds to a frequency stability of $\delta f/f \sim 1 \times 10^{-14}\, \tau^{-1/2}$, a value better than any frequency standard available at that time for measuring times equal to a typical two-way delay between Earth and Mars, and the 100 fW required power allows application at Mars' farthest distance from Earth with a reasonable transmitter power level. However, weak-light manipulation and weak-light technology still needs developments.

## 5. DISCUSSION AND OUTLOOK

GW astronomy has been an important drive to technology. Now it is growing. It will even be so. Let's discern what will come in next 50-100 years.

## ACKNOWLEDGEMENTS

We would like to thank Science and Technology Commission of Shanghai Municipality (STCSM-14140502500) and Ministry of Science and Technology of China (MOST-2013YQ150829, MOST-2016YFF0101900) for supporting this work in part.